\documentclass[conference]{IEEEtran}
\usepackage{times}

\usepackage[numbers]{natbib}
\usepackage{multicol}
\usepackage{graphicx}
\usepackage{subcaption}
\usepackage[bookmarks=true]{hyperref}
\usepackage{romannum}
\usepackage{mathtools}
\usepackage{amssymb}
\usepackage{amsmath}
\newcommand{\mytag}[1]{\tag*{(#1)}}
\newcommand{\HRule}{\noindent\rule{\linewidth}{0.1mm}\newline}
\newcommand{\argmin}{\mathop{\mathrm{argmin}}}
\usepackage{balance}

\IEEEoverridecommandlockouts
\begin{document}

\title{Reinforcement Learning for Safety-Critical Control under Model Uncertainty, using Control Lyapunov Functions and Control Barrier Functions}

\author{\authorblockN{Jason Choi$^{*1}$\thanks{$^*$ Indicates equal contribution.}, Fernando Casta\~neda$^{*1}$, Claire J. Tomlin$^{2}$, Koushil Sreenath$^{1}$}
\authorblockA{$^1$Department of Mechanical Engineering, $^2$Department of Electrical Engineering and Computer Sciences, UC Berkeley\\
Email: \{jason.choi, fcastaneda, tomlin, koushils\}@berkeley.edu}
\thanks{The work of Jason Choi received the support of a fellowship from Kwanjeong Educational Foundation, Korea. The work of Fernando Casta\~neda received the support of a fellowship (code LCF/BQ/AA17/11610009) from ”la Caixa” Foundation (ID 100010434). This work was partially supported through National Science Foundation Grant CMMI-1931853.}
}

\maketitle

\begin{abstract}
In this paper, the issue of model uncertainty in safety-critical control is addressed with a data-driven approach. For this purpose, we utilize the structure of an input-ouput linearization controller based on a nominal model along with a Control Barrier Function and Control Lyapunov Function based Quadratic Program (CBF-CLF-QP). Specifically, we propose a novel reinforcement learning framework which learns the model uncertainty present in the CBF and CLF constraints, as well as other control-affine dynamic constraints in the quadratic program. The trained policy is combined with the nominal model-based CBF-CLF-QP, resulting in the \emph{Reinforcement Learning-based CBF-CLF-QP (RL-CBF-CLF-QP)}, which addresses the problem of model uncertainty in the safety constraints. The performance of the proposed method is validated by testing it on an underactuated nonlinear bipedal robot walking on randomly spaced stepping stones with one step preview, obtaining stable and safe walking under model uncertainty.
\end{abstract}

\IEEEpeerreviewmaketitle

\section{Introduction}

In this work, we address the issue of model uncertainty in safety-critical control using a data-driven machine learning approach. Our goal is to benefit from the recent successes of learning-based control in highly uncertain dynamical systems, such as in \citet{hwangbo2019learning} and \citet{levine}, yet to also account for safety in a formal way.  We seek to combine the benefits of these data-driven approaches with the benefits of classical model-based control methods which have theoretical guarantees on stability and safety. Towards this end, we use Control Lyapunov Function- and Control Barrier Function-based controllers designed on nominal systems that are then trained through reinforcement learning (RL) to work on systems with uncertainty.

\subsection{Related Work}

In the field of controls, Control Lyapunov Function (CLF)-based and Control Barrier Function (CBF)-based control methods have been shown to be successful for safety-critical control. \citet{7079382} and \citet{ames2013towards} have shown that CLF-based quadratic programs (CLF-QP) with constraints can be solved online in order to perform locomotion and manipulation tasks. In \citet{ames2014cbf}, CBFs are incorporated with the CLF-QP, namely CBF-CLF-QP, to handle safety constraints effectively in real time.

These CLF-based and CBF-based methods heavily rely on accurate knowledge of the system model. When the model is uncertain, we must consider adaptive or robust versions. In \citet{nguyen2015L1}, an $L_1$ adaptive controller is incorporated with the CLF-QP in order to adapt to model uncertainty, and is shown to work effectively for bipedal walking. In \citet{nguyen2016optimal}, a robust version of the CBF-CLF-QP is proposed, that solves the quadratic program for the worst case effect of model uncertainty. While these methods can tackle model uncertainty to some degree, they may often fail to account for the correct magnitudes of adaptation and uncertainty.

Recently, several methods addressing the issue of model uncertainty in the control problem using a data-driven approach have been proposed. \citet{westenbroek2019feedback} proposes an RL-based method to learn the model uncertainty compensation for input-output linearization control. In \citet{castaneda2020} the former method is extended to underactuated bipedal walking on a flat terrain.
\citet{taylor2019episodic} and  \citet{taylor2019learning} each addresses how to learn the uncertainty in CLF and CBF constraints respectively, using empirical risk minimization. Our methodologies most closely align with these works in that we are also using learning methods to reduce model uncertainty explicitly in input-output linearization, CLF, and CBF-based control. However, the main novelty in our approach is that we have devised a unified RL-based framework for learning model uncertainty in CLF, CBF, and other dynamic control-affine constraints altogether in a single learning process. In addition to the aforementioned papers, there are also a few approaches \cite{bansal2017aDOBO,NIPS2017_6692,fisac} that learn model uncertainty through probabilistic models such as Gaussian Processes. Although these approaches allow for an insightful analysis of the learned model or policy, they can scale poorly with state dimension.

\begin{figure*}
\begin{center}
\includegraphics[width=17cm]{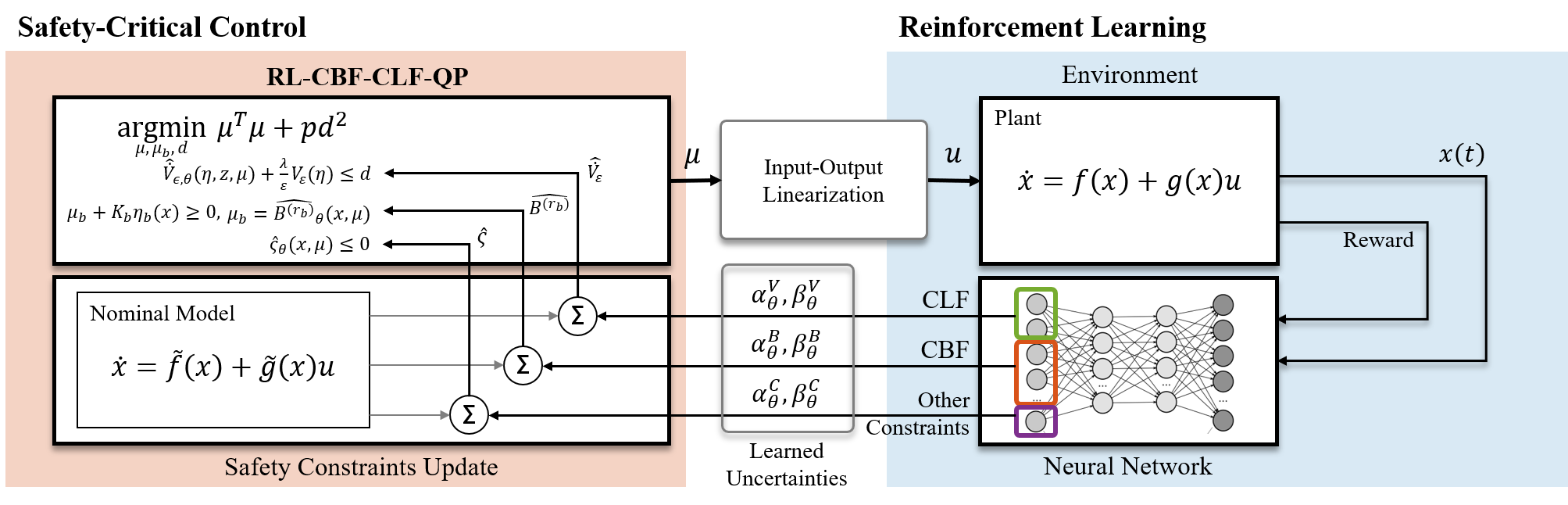}
\end{center}
\vspace{-1em}
\caption{Our method (RL-CBF-CLF-QP): We propose a control barrier function (CBF) and control Lyapunov function (CLF) based parametrized quadratic program, where the parameter $\theta$ corresponds to weights of a neural network that estimates the uncertainty in the CLF and CBF dynamics through reinforcement learning. An RL agent is used to learn uncertainties in the CLF, CBF and other constraint dynamics. The quadratic program uses learned uncertainties in combination with safety and stability constraints from a nominal model to solve for the control input point-wise in time.}\label{fig:Main}
\vspace{-1em}
\end{figure*}

\subsection{Contributions}
In this paper, we present a novel RL-based framework which combines two key components: 1) an RL agent which learns model uncertainty in multiple general dynamic constraints including CLF and CBF constraints through training, and 2) a quadratic program that solves for the control that satisfies the safety constraints under the learned model uncertainty. We name this framework \textbf{Reinforcement Learning-based Control Barrier Function and Control Lyapunov Function Quadratic Program (RL-CBF-CLF-QP)}. After training, the RL-CBF-CLF-QP can be executed online with fast computation. The overall diagram of our framework is presented in Fig. \ref{fig:Main}. Here is the summary of the contribution of our work:
\begin{enumerate}
    \item We present an RL framework that learns model uncertainty for CLF, CBF and other control-affine dynamic constraints in a single learning process.
    \item We generalize our method to high relative-degree outputs and Control Barrier Functions.
    \item Our method can learn the uncertainty in the dynamics of parameterized CBFs that are not only state-dependent but also dependent on other parameters.
    \item We numerically validate our method on an underactuated nonlinear hybrid system: a bipedal robot walking on stepping stones with significant model uncertainty.
\end{enumerate}

\subsection{Organization}
In Section \ref{sec:background}, we briefly explain the necessary background for the paper. In Section \ref{Sec-CLF-QP}, we discuss how we can learn model uncertainty in the CLF constraint for CLF-QP through RL. In Section \ref{sec:RL-CBF}, we expand this method to learn uncertainties in the CBF and general control-affine dynamic constraints, and propose the RL-CBF-CLF-QP. In Section \ref{SectionRL}, we discuss how the RL agent can learn aforementioned uncertainties. In Sections \ref{sec:application} and \ref{sec:results}, we explain the results of the demonstration of RL-CBF-CLF-QP for a bipedal robot. Finally, we discuss limitations of our method in Section \ref{sec:discussion} and give concluding remarks in Section \ref{sec:conclusion}.

\section{Background} 
\label{sec:background}
\subsection{Input-Output Linearization}
Consider a control affine nonlinear system
\begin{equation}
\label{system}
    \begin{aligned}
        &\dot{x}=f(x)+g(x)u,\\
        &y=h(x),
    \end{aligned}
\end{equation}
where $x \in \mathbb{R}^{n}$ is the system state, $u \in \mathbb{R}^m$ the control input and $y \in \mathbb{R}^m$ the output of the system, assuming there are the same number of input and output variables. We also make the standard assumption that $f$ and $g$ are Lipschitz continuous. Then, if the vector relative degree of the outputs is \emph{r}, we have

\begin{equation}
y^{(r)}=L_{f}^{r} h(x)+L_{g} L_{f}^{r-1} h(x) u, \label{eq:2}
\end{equation}

\noindent where the functions $L_{f}^{r}h$ and $L_{g}L^{r-1}_fh$ are known as $r^{th}$ order Lie derivatives \citep{lie}. Here, $y^{(r)}$ is the vector of $r^{th}$ derivatives of each output in $y$, and \eqref{eq:2} indicates that 
no input in $u$ appears at lower than the $r^{th}$ derivative of each output.
If the $m\times m$ matrix $L_{g}L^{r-1}_fh(x)$ is nonsingular $\forall$ $x\in D$, with $D\subset \mathbb{R}^{n}$ being a compact subset containing the origin, then we can apply a control input which renders the input-output dynamics of the system linear:
\begin{equation}
u(x, \mu)=u^{*}(x)+\left(L_{g} L_{f}^{r-1} h(x)\right)^{-1} \mu, \label{eq:3}
\end{equation}

\noindent where $u^{*}$ is the feedforward term:
\begin{equation}
u^{*}(x)=-\left(L_{g} L_{f}^{r-1} h(x)\right)^{-1} L_{f}^{r} h(x),
\end{equation}

\noindent and $\mu \in \mathbb{R}^m$ is the \textit{auxiliary input}.

Using this control law yields the input-output linearized system $y^{(r)} = \mu$, and we can define a state transformation $\Phi : x \rightarrow (\eta, z)$, with 
\begin{equation}
\eta =[h(x)^{\top},L_{f}h(x)^{\top},...,L_{f}^{r-1} h(x)^{\top}]^{\top} 
\end{equation}
and $z \in Z$, where $ Z=\{ x \in \mathbb{R}^{n} |\ \eta \equiv 0\}$ is the zero-dynamics manifold. The closed-loop dynamics of the system can then be represented as a linear time-invariant system on the transverse coordinates $\eta$, and the zero-dynamics:

 \begin{equation}
 \left\lbrace
      \begin{aligned}
         \dot{\eta} & = F\eta +G\mu,\\
         \dot{z} & = p(\eta,z),
      \end{aligned}
    \right.
    \label{eq:linear-zero-dynamics}
\end{equation}
where
\begin{equation}
F=\left[\begin{array}{ccccc}
{0} & {I_m} & {.} & {.} & {0} \\
{0} & {0} & {I_m} & {.} & {0} \\
{.} & {.} & {.} & {} & {} \\
{0} & {.} & {.} & {.} & {I_m} \\
{0} & {.} & {.} & {.} & {0}
\end{array}\right] \text { and } G=\left[\begin{array}{l}
{0} \\
{.} \\
{.} \\
{0} \\
{I_m}
\end{array}\right],
\end{equation}
with $F\in \mathbb{R}^{mr\times mr}$ and $G\in \mathbb{R}^{mr\times m}$.

\subsection{Control Lyapunov Function Based Quadratic Programs}
\label{subsec:CLF-QP}

In \citet{6709752} a control method that guarantees exponential stability of the transverse dynamics $\eta$ with a rapid enough convergence rate is presented. It introduces the concept of a \emph{rapidly exponentially stabilizing control Lyapunov function (RES-CLF)}. Specifically, a one-parameter family of continuously differentiable functions $ V_{\varepsilon} : \mathbb{R}^{mr} \rightarrow \mathbb{R} $ is said to be an RES-CLF for system \eqref{system} if $\exists$ $\gamma$, $c_1$, $c_2 > 0$ such that $\forall$ $0 < \varepsilon < 1$ and $\forall$ $\eta \in \mathbb{R}^{mr}$, the following holds:
\begin{equation}
c_{1}\|\eta\|^{2} \leq V_{\varepsilon}(\eta) \leq \frac{c_{2}}{\varepsilon^2}\|\eta\|^{2},
\end{equation}
\vspace{-13pt}
\begin{equation}
\dot{V_\varepsilon}(\eta, \mu)+\frac{\lambda}{\varepsilon} V_\varepsilon(\eta) \leq 0.
\label{CLF_cond}
\end{equation}

If we define a control input $\mu$ that makes $\eta$ exponentially stable, of the form
\begin{equation}
\mu = \left[-\frac{1}{\varepsilon^r} K_r,\ ..., \  -\frac{1}{\varepsilon^2} K_2, \  -\frac{1}{\varepsilon} K_1\right]\eta = K\eta,
\end{equation}
\noindent where $K\in\mathbb{R}^{m\times mr}$, then we can choose a quadratic CLF candidate $V_\varepsilon(\eta)=\eta^{T} P_\varepsilon \eta$, where $P_\varepsilon$ is the solution of the Lyapunov equation $A^{T}P_\varepsilon + P_\varepsilon A = -Q$, with $A$ being the closed-loop dynamics matrix $A = F+GK$ and Q any symmetric positive-definite matrix. Defining $\bar{f}=F \eta$, $\bar{g}=G$, we can write the derivative of the RES-CLF as:
\begin{equation}
\dot{V_\varepsilon}(\eta, \mu)=L_{\bar{f}} V_\varepsilon(\eta)+L_{\bar{g}} V_\varepsilon(\eta) \mu,
\end{equation}
with 
{\small 
\begin{equation}
\begin{aligned}
&L_{\bar{f}} V_\varepsilon(\eta)=\eta^{T}\left(F^{T} P_\varepsilon+P_\varepsilon F\right) \eta, \quad L_{\bar{g}} V_\varepsilon(\eta)=2 \eta^{T} P_\varepsilon G.
\label{eq:clf-lie}
\end{aligned}
\end{equation}
}

We can then define for every time step an optimization problem in which condition \eqref{CLF_cond} becomes a linear constraint on the auxiliary input $\mu$. The objective function can be set to minimize the norm of the control inputs, in which case the optimization problem is a quadratic program (QP):\\

\hrule \vspace{5pt}
\textbf{CLF-QP}:
\begin{align}
    \label{CLF_QP}
    \mu^{*}(x)=\underset{\mu}{\operatorname{argmin}} \ & \mu^{T} \mu \\
    \tag{CLF}
        \text { s.t. } \ & \dot{V_\varepsilon}(\eta, \mu)+\frac{\lambda}{\varepsilon} V_\varepsilon(\eta) \leq 0
\end{align}

\subsection{Control Barrier Function and Control Lyapunov Function Based Quadratic Programs}
\label{subsec:CBF-CLF}
In \citet{7524935} the concept of an Exponential Control Barrier Function (ECBF) is defined. Specifically, a function $B:\mathbb{R}^m\rightarrow\mathbb{R}$ is an ECBF of relative degree $r_b$ for the system \eqref{system} if there exists $K_b\in\mathbb{R}^{1\times r_b}$ such that 
\begin{equation}
\sup _{u}\left[L_{f}^{r_{b}} B(x)+L_{g} L_{f}^{r_{b}-1} B(x) u+K_{b} \eta_{b}(x)\right] \geq 0
\label{CBF_const}
\end{equation}
for $\forall x \in \lbrace x \in \mathbb{R}^n |\ B(x)\geq 0\rbrace$ 
with 
{\small 
\begin{equation}
\eta_{b}(x)=\left[\begin{array}{c}
{B(x)} \\
{\dot{B}(x)} \\
{\ddot{B}(x)} \\
{\vdots} \\
{B^{\left(r_{b}-1\right)}(x)}
\end{array}\right]=\left[\begin{array}{c}
{B(x)} \\
{L_{f} B(x)} \\
{L_{f}^{2} B(x)} \\
{\vdots} \\
{L_{f}^{r_{b}-1}}{B(x)}
\end{array}\right],
\end{equation}
}
that guarantees $B(x_0)\geq0 \implies B(x(t)) \geq0,\ \forall t \geq0$.

We can then choose a \textit{virtual input} $\mu_b$ that input-output linearizes the ECBF dynamics:
\begin{equation}
B^{(r_{b})}(x,\mu) = L^{r_b}_{f}B(x) + L_{g}L^{r_{b}-1}_{f} B(x) u(x,\mu) =:\mu_{b},
    \label{eq:cbf_der}
\end{equation}
with $u$ defined in \eqref{eq:3}.  We refer readers to \citet{7524935} for more details. The condition in \eqref{CBF_const} then translates to choosing a $\mu_b$ such that
\begin{equation}
    \mu_b + K_b\eta_b\geq 0,
\end{equation}
which is added to the following QP, where safety is prioritized over stability by relaxing the CLF constraint:

\HRule
\noindent \textbf{CBF-CLF-QP}:
\begin{align}
\label{CBF-QP}
\mu^{*}(x)=	\ & \underset{\mu,\ \mu_b,\  d}{\argmin} & & \mu^T \mu + p~d^2 \\
& \text{s.t.} & & \dot{V_\varepsilon}(\eta, \mu)+\frac{\lambda}{\varepsilon} V_\varepsilon(\eta) \leq d  \mytag{CLF} \\
& & &\mu_b +K_b\ \eta_b \geq 0  \mytag{CBF} \\
&&& \mu_b = B^{(r_b)}(x,\mu) \notag \\
&&& A_{c}(x) \mu  + b_{c}(x) \leq 0 \mytag{Constraints}
\end{align}

Formulating a QP allows us to incorporate additional control-affine constraints (last line in \eqref{CBF-QP}). These could be input saturation constraints or other state-dependent constraints such as contact-force constraints.

\section{Reinforcement Learning for CLF-QP Based Controllers under Uncertain Dynamics}
\label{Sec-CLF-QP}

In this section, we address the issue of having a mismatch between the model and the plant dynamics when the true plant vector fields $f,\ g$ are not precisely known. Specifically, between this and the next sections we analytically examine the effects of model uncertainty on the dynamics of the CLF, CBF and other control-affine dynamic constraints. For each of these cases we will define the goal of the RL agent and the policy to be learned.

\subsection{Reinforcement Learning for CLF-QP Based Controllers: First Approach}
\label{Approach1}

Let the \textit{nominal model} used in the controller be
 \begin{equation}
     \dot{x} = \tilde{f}(x) + \tilde{g}(x)u.
 \end{equation}
 
We assume: 1) the vector fields $\tilde{f}:\mathbb{R}^{n}\rightarrow\mathbb{R}^{n}, \tilde{g}:\mathbb{R}^{n}\rightarrow\mathbb{R}^{n\times m}$ are Lipschitz continuous and 2) the vector relative degrees of the model and plant dynamics are the same ($r$). These are the standard assumptions that have been made in most of the literature \cite{nguyen2016optimal,taylor2019episodic,taylor2019learning,westenbroek2019feedback} to tackle the mismatch terms analytically.
 
The pre-control law \eqref{eq:3} of input-output linearization computed based on the nominal model $\tilde{f}, \tilde{g}$ has the following form
 \begin{equation}
    \tilde{u}(x, \mu) = \tilde{u}^{*}(x) + \left(L_{\tilde{g}} L_{\tilde{f}}^{r-1}h(x)\right)^{-1} \mu,
    \label{eq:io_input_model}
\end{equation}
with a feedforward term
\begin{equation}
    \tilde{u}^{*}(x) \coloneqq -\left(L_{\tilde{g}} L_{\tilde{f}}^{r-1}h(x)\right)^{-1} L^{r}_{\tilde{f}} h(x).
\end{equation}
 Using this $\tilde{u}$ in \eqref{eq:2} yields
\begin{equation}
    y^{(r)} = \mu + \mathit{\Delta_{1}(x)} + \mathit{\Delta_{2}(x)} \mu,
\label{uncert}
\end{equation}
where
{\small
\begin{equation}
    \begin{aligned}
    \mathit{\Delta_{1}(x)} \coloneqq& L^{r}_{f} h(x) - L_{g} L_{f}^{r-1}h(x)\left(L_{\tilde{g}} L_{\tilde{f}}^{r-1}h(x)\right)^{-1} L^{r}_{\tilde{f}} h(x), \\
    \mathit{\Delta_{2}(x)} \coloneqq& L_{g} L_{f}^{r-1}h(x)\left(L_{\tilde{g}} L_{\tilde{f}}^{r-1}h(x)\right)^{-1} - \mathit{I}_m.
\end{aligned}
\label{delta_IO2}
\end{equation}
}
The dynamics of $\eta$ from \eqref{eq:linear-zero-dynamics} now take the form:
\begin{equation}
    \dot{\eta} = \left(F\eta + G\mathit{\Delta_{1}(\eta, z)}\right) +  G\left(I_m + \mathit{\Delta_{2}(\eta, z)} \right) \mu.
    \label{eq:nonlinear-zero-dynamics}
\end{equation}

Note that this equation is the same as \eqref{eq:linear-zero-dynamics} if the uncertainty terms are zero, i.e. $\mathit{\Delta_{1}}=\mathit{\Delta_{2}}=0$. Thus, \eqref{eq:linear-zero-dynamics} can be considered a nominal model for the true transverse dynamics \eqref{eq:nonlinear-zero-dynamics}.

For this first approach we use RL to define an additional input whose goal is to cancel out the uncertainty terms present in the transverse dynamics \eqref{eq:nonlinear-zero-dynamics}, and therefore manipulate the transverse dynamics to behave like \eqref{eq:linear-zero-dynamics}, as done in \citet{castaneda2020} and \citet{westenbroek2019feedback}. If this is achieved exactly, there will not be any uncertain terms in the CLF dynamics, since $\dot{V}_\varepsilon$ only depends on the matrices $F$ and $G$ of the input-output linearized dynamics.

Applying the following input to \eqref{eq:2}
\begin{equation}
    u(x,\mu) = \tilde{u}(x,\mu) + u_\theta(x,\mu),
    \label{approach1_input}
\end{equation}
with $\tilde{u}$ as defined in \eqref{eq:io_input_model} and with
\begin{equation}
    u_\theta(x,\mu) \coloneqq \left(L_{\tilde{g}} L_{\tilde{f}}^{r-1}h(x)\right)^{-1} (\alpha_\theta(x)\mu + \beta_\theta(x)),
\end{equation}
yields
\begin{small}
\begin{equation}
    y^{(r)} = \mu + \left(\mathit{\Delta_{1}(x)} + \mathit{\Delta_{3}(x)}\beta_\theta(x)\right) + \left(\mathit{\Delta_{2}(x)}+\mathit{\Delta_{3}(x)}\alpha_\theta(x)\right) \mu,
\vspace{-5pt}
\end{equation}
\end{small}

\noindent where $\mathit{\Delta_{3}(x)} \coloneqq \mathit{\Delta_{2}(x)} + I_m$, and $\theta \in \Theta \subset \mathbb{R}^{N}$ are parameters of a neural network to be learned.
We can now clearly see the goal of the RL agent for this approach: design a policy $\alpha_\theta$, $\beta_\theta$ such that $y^{(r)}$ is as close as possible to $\mu$. Thus, the time-wise reward function can be defined as 
\begin{equation}
    R(x,\mu) = -||y^{(r)}-\mu||_2^2\,
\end{equation}
where $y^{(r)}$ is numerically estimated. After training, the $\mu$ present in the final control input \eqref{approach1_input} is obtained by solving the CLF-QP of \eqref{CLF_QP} in real time. We call this first approach \textit{IO-RL + CLF-QP}.

\subsection{Reinforcement Learning for CLF-QP Based Controllers: Second Approach}
\label{Approach2}

In the second approach, we do not directly correct the uncertain terms of the transverse dynamics \eqref{eq:nonlinear-zero-dynamics} as we did in the first approach. Instead, we directly analyze the impact of this uncertainty on the dynamics of the CLF. 

For this approach, we assume that the CLF designed for the nominal model's transverse dynamics is also a CLF for the true plant's transverse dynamics \eqref{eq:nonlinear-zero-dynamics}.

In the presence of uncertainty, $\dot{V}_\varepsilon$ becomes
\begin{equation}
    \dot{V}_{\varepsilon}(\eta,z,\mu) = L_{\bar{f}}V_{\varepsilon}(\eta,z)+  L_{\bar{g}}V_{\varepsilon}(\eta,z) \mu,
\end{equation}
where 
\begin{equation}
\begin{aligned}
   L_{\bar{f}}V_{\varepsilon}(\eta,z) &= L_{\tilde{\bar{f}}}V_{\varepsilon}(\eta) + \underbrace{2 \eta^{\intercal} P_{\varepsilon} G \mathit{\Delta_{1}(\eta, z)}}_\text{$=:\mathit{\Delta^v_{1}(\eta,z)}$}, \\
   L_{\bar{g}}V_{\varepsilon}(\eta,z) &= L_{\tilde{\bar{g}}}V_{\varepsilon}(\eta) + \underbrace{2 \eta^{\intercal} P_{\varepsilon}G \mathit{\Delta_{2}(\eta, z)}}_\text{$=:\mathit{\Delta^v_{2}(\eta,z)}$}.
\end{aligned}
\label{eq70}
\end{equation}
Here, $\tilde{\bar{f}}$ and $\tilde{\bar{g}}$ are the nominal model input-output linearized dynamics: namely, $\widetilde{\dot{V}}_{\varepsilon}(\eta, \mu) = L_{\tilde{\bar{f}}}V_{\varepsilon}(\eta) + L_{\tilde{\bar{g}}}V_{\varepsilon}(\eta) \mu$.
Therefore, under uncertainty:
\begin{equation}
    \dot{V}_{\varepsilon}(\eta,z, \mu) = \widetilde{\dot{V}}_{\varepsilon}(\eta, \mu) + \mathit{\Delta^v_{1}(\eta,z)}+\mathit{\Delta^v_{2}(\eta,z)} \mu.
\end{equation}

In this second approach we use RL to estimate the uncertainty terms in $\dot{V}_\varepsilon$: $\mathit{\Delta^v_{1}}$ and $\mathit{\Delta^v_{2}}$. For this purpose, we construct an estimate
\begin{equation}
\label{CLF_estimate}
     \widehat{\dot{V}}_{\varepsilon, \theta}(\eta,z, \mu) = \widetilde{\dot{V}}_{\varepsilon}(\eta, \mu) + \beta^{V}_{\theta}(\eta,z) + \alpha^{V}_{\theta}(\eta,z)\mu,
\end{equation}
where $\theta \in \Theta \subset \mathbb{R}^{N}$ are again the neural network parameters to be learned. The goal of RL is then obvious: learn a policy $\alpha^V_\theta$, $\beta^V_\theta$ such that $\widehat{\dot{V}}_{\varepsilon, \theta}$ is as close as possible to $\dot{V}_{\varepsilon}$. Any reward function that penalizes the absolute value of the difference between the two terms can be used. More details on the specific RL implementation are discussed in Section \ref{SectionRL}.

\smallbreak

\textbf{Remark 1}: For convenience, it is assumed here that $\alpha_\theta^{V}$, $\beta_\theta^{V}$ share the same network parameters $\theta$, but this does not need to be the case. In this paper, we will assume that all the policy functions to be learned are sharing the same parameters.
\smallbreak

The estimate $\hat{\dot{V}}_{\varepsilon,\theta}$ in \eqref{CLF_estimate} is then used as our best guess of $\dot{V}_\varepsilon$ for the optimization problem:

\HRule
\noindent \textbf{RL-CLF-QP}:
\begin{align}
\label{RL-CLF-QP}
\mu^{*}_\theta(x)=	\ & \underset{\mu}{\argmin} & & \mu^T \mu\\
& \text{s.t.} & & \widehat{\dot{V}}_{\varepsilon, \theta}(\eta,z, \mu)+\frac{\lambda}{\varepsilon} V_\varepsilon(\eta) \leq 0  \mytag{RL-CLF}
\end{align}

\textbf{Remark 2}: We have illustrated the case in which the CLF is applied to the input-output linearized dynamics. The reason why we use a CLF on the input-output linearized dynamics instead of the full dynamics is that in this way we have a systematic way of computing a CLF candidate, whereas on the original nonlinear system this process could be challenging. However, this approach is not confined to the input-output linearization structure and is also applicable to any general nonlinear control-affine system.

\section{Reinforcement Learning for CBF-CLF-QP Based Controllers under Uncertain Dynamics}

Having studied how to compensate for the effects of model uncertainty on CLF-based min-norm controllers, we will now extend our framework to the safety-critical CBF-CLF-QP by following a similar approach.

\label{sec:RL-CBF}
\subsection{Reinforcement Learning for CBFs}
In the presence of uncertainty, \eqref{eq:cbf_der} becomes
\begin{equation}
    \widetilde{B^{(r_b)}}(x, \mu) = L^{r_b}_{\tilde{f}}B(x) + L_{\tilde{g}}L^{r_{b}-1}_{\tilde{f}} B(x) \tilde{u}(x, \mu),
    \label{eq80}
\end{equation}
and the actual CBF's $r_b^{th}$ derivative can be written as:
\begin{equation}
    B^{(r_b)}(x, \mu) = \widetilde{B^{(r_b)}}(x, \mu) + \mathit{\Delta^{b}_{1}}(x) + \mathit{\Delta^{b}_{2}}(x) \mu,
\end{equation}
where $\mathit{\Delta^{b}_{1}}$ and $\mathit{\Delta^{b}_{2}}$ are the uncertain terms that arise from the model-plant mismatch. We omit analytic expressions of $\mathit{\Delta^{b}_{1}}, \mathit{\Delta^{b}_{2}}$ for conciseness, but they can be derived similarly to \eqref{delta_IO2}.

\textbf{Remark 3}: When the state of the system can be represented as $x=[q,\dot{q}]^T$, as in most robotic systems, even for high relative degree CBFs model uncertainty only affects the $r^{th}_{b}$ time derivative of $B$, since $B^{(r_{b})}$ is the only term that depends on the plant dynamics through the vector fields $f$ and $g$. 
\smallbreak

Next, we present how to estimate the uncertainty terms for the CBF and for other dynamic constraints using RL. The approach presented in Section \ref{Approach1} cannot be used here since the CBF functions depend on the full dynamics of the system, and not the transverse dynamics.

We build an estimator of $B^{(r_b)}$:
\begin{equation}
     \widehat{B^{(r_b)}}_{\theta}(x, \mu) = \widetilde{B^{(r_b)}}(x, \mu) + \beta^{B}_{\theta}(x) + \alpha^{B}_{\theta}(x) \mu,
\end{equation}
and the goal of RL is to learn a policy $\alpha^{B}_{\theta}$, $\beta^{B}_{\theta}$ such that $\widehat{B^{(r_b)}}_{\theta}$ is as close as possible to $B^{(r_b)}$.

In order to integrate everything in a new QP we define the new virtual input of the CBF dynamics as
\begin{equation}
    \mu_b \coloneqq \widehat{B^{(r_b)}}_{\theta}.
\end{equation}

In cases where the CBF also depends on a set of parameters $\psi \in \mathbb{R}^q$ 
, then we need to define the CBF as $B:\mathbb{R}^{n\times q} \rightarrow \mathbb{R}$. The neural-network policy will now need to take $\psi$ as additional inputs $\alpha^{B}_{\theta}:\mathbb{R}^{n\times q}\rightarrow\mathbb{R}^{m}, \ \beta^{B}_{\theta}:\mathbb{R}^{n\times q}\rightarrow\mathbb{R}$ and the proposed estimate of the $r_b^{th}$ time derivative of $B$ becomes:
\begin{equation}
     \widehat{B^{(r_b)}}_{\theta}(x, \mu, \psi) = \widetilde{B^{(r_b)}}(x, \mu, \psi)+ \beta^{B}_{\theta}(x,\psi) + \alpha^{B}_{\theta}(x,\psi) \mu.
\end{equation}

\subsection{Reinforcement Learning for Additional Control-Affine Dynamic Constraints}
\label{subsec:RL-CBF-linear-constraint}
Now we study the effects of uncertainty on other linear constraints that depend on the dynamics of the system: 
\begin{equation}
    \underbrace{A_{c}(x,f,g) \mu  + b_{c}(x,f,g)}_\text{$=:\zeta(x,\mu)$} \leq 0.
\end{equation}
In the presence of model mismatch we have
\begin{equation}
\begin{aligned}
  b_{c}(x,f,g) &= b_{c}(x,\tilde{f},\tilde{g}) + \mathit{\Delta^c_{1}(x)}, \\
  A_{c}(x,f,g) &= A_{c}(x,\tilde{f},\tilde{g}) + \mathit{\Delta^c_{2}(x)},
\end{aligned}
\label{eq_const}
\end{equation}
where $\mathit{\Delta^c_{1}}$ and $\mathit{\Delta^c_{2}}$ represent the uncertainty terms. We can then define the nominal constraint
\begin{equation}
    \tilde{\zeta}(x,\mu) = b_{c}(x,\tilde{f},\tilde{g}) + A_{c}(x,\tilde{f},\tilde{g}) \mu.
\end{equation}

\noindent And the real value of the constraint can be expressed as
\begin{equation}
    \zeta(x,\mu) = \tilde{\zeta}(x,\mu) + \mathit{\Delta^c_{1}(x)} +  \mathit{\Delta^c_{2}(x)} \mu.
\end{equation}

\noindent We can build an estimator of the form
\begin{equation}
    \hat{\zeta}_\theta (x,\mu) = \tilde{\zeta}(x,\mu) + \beta^C_\theta (x) + \alpha^C_\theta (x) \mu,
\end{equation}
with a learned policy $\alpha^C_\theta$, $\beta^C_\theta$. The goal of the RL agent is again in this case to make the estimator  $\hat{\zeta}_\theta$ as close as possible to $\zeta$. Expanding $\tilde{\zeta}$ we can rewrite the estimator as
\begin{equation}
    \hat{\zeta}_\theta (x,\mu) = \underbrace{\left(b_{c}(x,\tilde{f},\tilde{g}) + \beta^C_\theta (x)\right)}_\text{$=:b^c_\theta (x)$} + \underbrace{\left(A_{c}(x,\tilde{f},\tilde{g}) + \alpha^C_\theta (x) \right)}_\text{$=:A^c_\theta (x)$} \mu.
\end{equation}

So far, we have explained our method of constructing an estimator of a single $B^{(r_b)}$ and a single $\zeta(x,\mu)$. This can be applied to $n_b$ multiple CBFs and $n_c$ multiple control-affine constraints. The final optimization problem, which includes all the learned estimates of the uncertain terms is:

\HRule
\noindent \textbf{RL-CBF-CLF-QP}:
\begin{align}
\label{RL-CBF-CLF-QP}
\mu^{*}_\theta(x)= &\underset{\mu,\ \mu_b,\  d}{\argmin} && \mu^T \mu + p~d^2 \\
& \text{s.t.} && \widehat{\dot{V}}_{\varepsilon, \theta}(\eta,z, \mu)+\frac{\lambda}{\varepsilon} V_\varepsilon(\eta) \leq d  \mytag{RL-CLF} \\
\text {for } i=1\cdots n_{b} \hspace{-30pt} &&&\mu_{b,i} +K_{b,i}\ \eta_{b,i} \geq 0  \mytag{RL-CBF} \\
&&&\mu_{b, i} = \widehat{B^{(r_b)}}_{i, \theta}(x, \mu) \notag \\
\text {for } j=1\cdots n_{c} \hspace{-30pt} &&&A^c_{j,\theta} (x) \mu + b^c_{j,\theta} (x) \leq 0 \mytag{RL-Constraints}
\end{align}

\section{Reinforcement Learning-based Framework}
\label{SectionRL}

In this section, we present a unified RL framework that can learn the uncertainty terms in the CLF, CBF, and other dynamic constraints by building the terms specified in the earlier sections as $\alpha_\theta^V, \alpha_\theta^B, \alpha_\theta^C, \beta_\theta^V, \beta_\theta^B, \beta_\theta^C$.

A diagram of this framework is illustrated in Fig. \ref{fig:Main}. The RL agent learns a policy, which is a combination of uncertainty terms in CLF, CBF and other dynamic constraints. These terms are then added to the QP constraints derived from the nominal model, resulting in the estimates of the true plant constraints. Using these estimates, the RL-CBF-CLF-QP optimization problem, in which model uncertainty is addressed, is solved point-wise in time to obtain the control input. 

The reward function of the learning problem is designed such that it minimizes each of the estimation errors. Thus, the time-wise loss functions are defined as:

\begin{equation}
\label{eq:loss_functions}
    \begin{aligned}
    l_{V, \theta} :&= ||\dot{V}_{\varepsilon} - \widehat{\dot{V}}_{\varepsilon, \theta}(x, \mu)||^2 \\
    l_{B, \theta} :&= ||B^{(r_b)} - \widehat{B^{(r_b)}}_{\theta}(x, \mu)||^2 \\
    l_{C, \theta} :&= ||\zeta - \hat{\zeta}_\theta(x,\mu)||^2
    \end{aligned}
\end{equation}

It is important to note that the true plant's dynamics information is not used for computing the values of these loss functions. We use explicit expressions for $V_\varepsilon$, $B$ and $\zeta$ and compute the time-derivatives
$\dot{V}_{\varepsilon}$, $B^{(r_b)}$ using numerical differentiation. For the CBF, it is important to note that regardless of the value of $r_b$ we only need to do numerical differentiation once, as follows from Remark 3.

A canonical RL problem can be formulated, with the reward for a given state $x$ defined as the weighted sum of the negative loss functions in \eqref{eq:loss_functions}, in addition to a user-specific failure-case penalty $-l_{e}: \mathbb{R}^n \rightarrow \mathbb{R}$:
\begin{equation}
    R(x, \theta) = -w_{v} l_{V, \theta} - \sum_{i=1}^{n_{b}} w_{b, i} l_{B_{i}, \theta} -  \sum_{j=1}^{n_{c}} w_{c, j} l_{C_{j}, \theta}  -l_{e}(x).
\end{equation}

The learning problem is then defined as:
\vspace{-5pt}
\begin{equation}
    \begin{aligned}
    \max _{\theta}\ & \mathbb{E}_{x_{0} \sim X_{0}, w \sim \mathcal{N}\left(0, \sigma^{2}\right)} \int_{0}^{T} R\left(x(\tau), \theta\right) d \tau, \\
    \text { s.t. } & \dot{x} = f(x) + g(x) \tilde{u}(x, \mu^{*}_{\theta}(x)+\omega),
    \end{aligned}
\end{equation}
where $\mu^{*}_{\theta}(x)$ is the solution of
\eqref{RL-CBF-CLF-QP}, $X_{0}$ is the initial state distribution, and $w \sim \mathcal{N}(0, \sigma^{2})$ is white noise added to encourage exploration. A discretized version of this problem can be solved using conventional RL algorithms.
\smallbreak

\textbf{Remark 4}: While running training experiments or simulations, it is assumed that the robot operates under the true plant dynamics.
We will later show in Section \ref{sec:results} that the trained policy works well even when the true plant in the evaluation differs from the plant of the training environment.

\section{Application to Bipedal Robots}
\label{sec:application}
The goal of this section is to validate that the RL-CBF-CLF-QP framework enables safety-critical control when model uncertainty is present. We test our method on RABBIT \cite{chevallereau2003rabbit}, a planar five-link bipedal robot, walking on a discrete terrain of stepping stones with one step preview.

\subsection{Simulation Settings}

We run two simulation scenarios with our method and offer comparisons with the previous methods. The first simulation consists of RABBIT simply walking on a flat terrain. We evaluate the CLF based methods in Section \ref{Sec-CLF-QP} in this scenario. This is to verify only the stabilizing capacity of our proposed method under model uncertainty. In the second simulation, we put the robot on a discrete terrain of randomly spaced stepping stones (Fig. \ref{fig:stepping-stone}). The robot's task here is to always place the foot on the next stepping stone, while managing the stability and not violating the contact-force constraint. The full RL-CBF-CLF-QP is tested in this simulation scenario.

The main model uncertainty in both demonstrations is introduced by scaling all mass and inertia parameters of each link by a constant scale factor = 2, i.e. the nominal model's mass and inertia terms are half of those of the actual plant.

A single periodic walking gait trajectory is generated offline by the Fast Robot Optimization and Simulation Toolkit (FROST) \citep{hereid2017frost}. The output function $h(x)$ is defined as the difference between the actuated joint angles and the desired trajectory's joint angles from the obtained periodic orbit. The gait's nominal step length is 0.35m. Finally, a torque saturation of 200Nm is applied to the control inputs of all simulations, including training and evaluation.

\subsection{Reinforcement Learning Settings}

We train our agent using a standard Deep Deterministic Policy Gradient Algorithm (DDPG) \citep{silver2014deterministic}.
The input for the actor neural network is 14 observations, which is RABBIT's full state $x$, in addition to the CBF parameter $\psi = l_{min, k}$ corresponding to the minimum step length of the $k$th stepping stone (Fig.~\ref{fig:stepping-stone}) in the second simulation. 
We use two CBFs $B_1$ and $B_2$ to constrain the position of the swing foot so that it lands on the stepping stone, as shown in Fig.~\ref{fig:stepping-stone}. We use two dynamic constraints $C_1$ and $C_2$ which correspond to the unilateral normal force and friction cone constraints respectively. The output dimension is 25, corresponding to the 4$\times$1 $\alpha^V_\theta, \alpha^{B_1}_{\theta}, \alpha^{B_2}_{\theta}, \alpha^{C_1}_{\theta}, \alpha^{C_2}_{\theta}$ and the 1$\times$1 $\beta^V_\theta, \beta^{B_1}_{\theta}, \beta^{B_2}_{\theta}, \beta^{C_1}_{\theta}, 
\beta^{C_2}_{\theta}$. 

Both actor and critic neural networks have two hidden layers of widths 400 and 300. This agent is trained on the simulation of ten walking steps per episode, and a discrete time step $T_s = 0.01$sec is used. The failure cases are determined by the robot's pose. Training on six multiple cores of Intel(R) Core(TM) i5-9400F CPU (2.90GHz) without the use of GPU took about 34 seconds per episode. The final agent in use is obtained after 110, 79 and 133 episodes for IO-RL + CLF-QP, RL-CLF-QP and RL-CBF-CLF-QP respectively.

\section{Results}
\label{sec:results}
During the evaluation, the robot is tested not only on the uncertainty that is introduced in the training, but in addition to it, two other kinds of uncertainty are also introduced. First, the robot's motor dynamics that restricts the rate of change of joint torques is applied in every evaluation. The time constant of motors used in the simulation is 0.004 seconds. Second, the robot is also tested on an alternative kind of uncertainty, which consists of an added weight to the torso of the robot, instead of scaling the links masses and inertias. This weight can represent the robot carrying a payload, and it is deliberately introduced to evaluate the trained policy's robustness to an unfamiliar kind of uncertainty that it was not trained on. 

\subsection{Simulation 1: Bipedal Walking on Flat Ground}

For the first simulation, we evaluate the two RL approaches for CLF explained in Section \ref{Sec-CLF-QP}, and compare them with the standard L1 Adaptive CLF-QP method of \citet{nguyen2015L1}, which guarantees the CLF to be bounded to a small value under model uncertainty if using a sufficiently large adaptation gain. 

\begin{figure}
\centering
\includegraphics[width=0.5\textwidth]{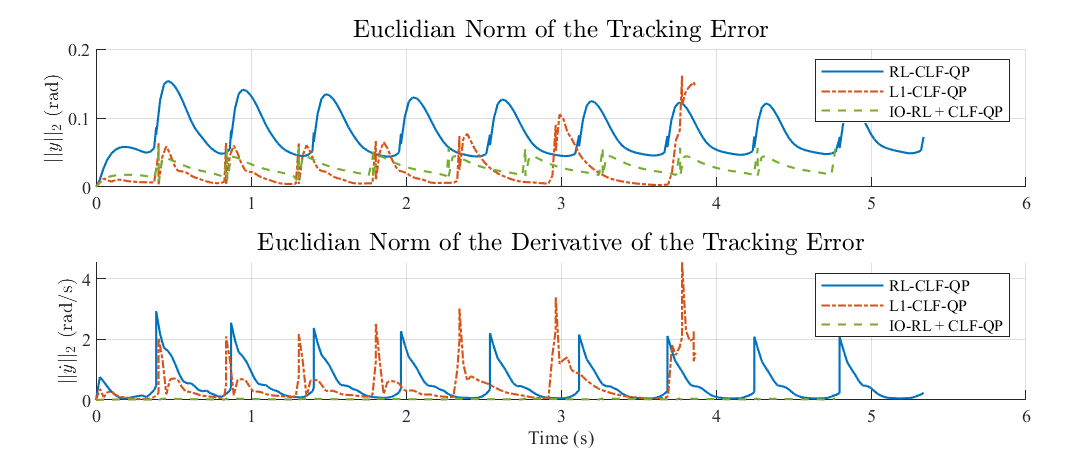}
\includegraphics[width=0.5\textwidth]{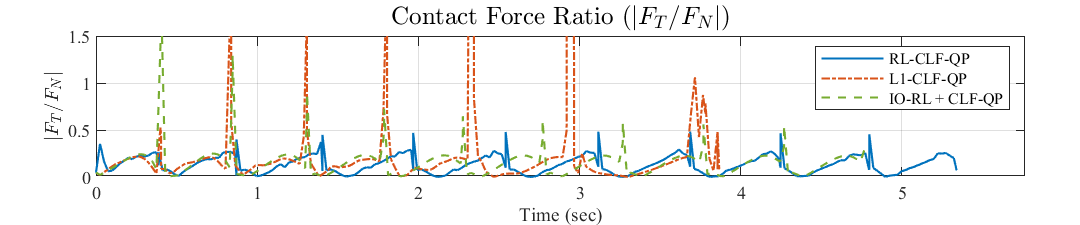}
\caption{Tracking error (top), its derivative (middle), and tangential-normal contact force ratio (bottom) of IO-RL + CLF-QP (Sec. \ref{Approach1}), RL-CLF-QP (Sec. \ref{Approach2}), and L1-CLF-QP \cite{nguyen2015L1} controllers, simulated for ten walking steps, where the robot's mass and inertia values are scaled by a factor of 2. Both IO-RL + CLF-QP and RL-CLF-QP maintain the stability while L1-CLF-QP fails. Only the RL-CLF-QP satisfies the friction cone constraint $|F_{T}/F_{N}|\leq k_{f} = 0.8$.}
\label{fig:clf-results}
\vspace{-10pt}
\end{figure}

As illustrated in Fig. \ref{fig:clf-results}, both of the proposed methods manage to get RABBIT to stably walk for multiple steps, while the L1 Adaptive CLF-QP controller leads to failure. The original nominal CLF-QP, although not shown in the figure, also fails under this scaled model uncertainty. Note that all three methods do not have friction constraints in the QP and could potentially violate them. In particular, the RL-CLF-QP method succeeds in satisfying the friction constraint ($|F_{T}/F_{N}|\leq k_{f} = 0.8$) for all steps, the IO-RL + CLF-QP exceeds the limit in the first two steps, and the L1-CLF-QP violates it for multiple steps. Therefore, IO-RL + CLF-QP needs the inclusion of friction constraints in the QP.

\begin{figure}
\centering
\includegraphics[width=0.5\textwidth]{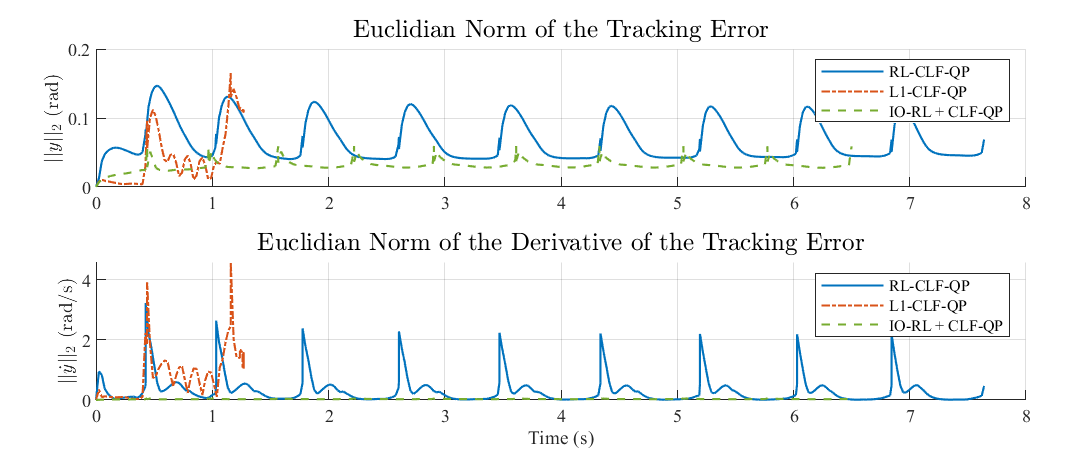}
\includegraphics[width=0.5\textwidth]{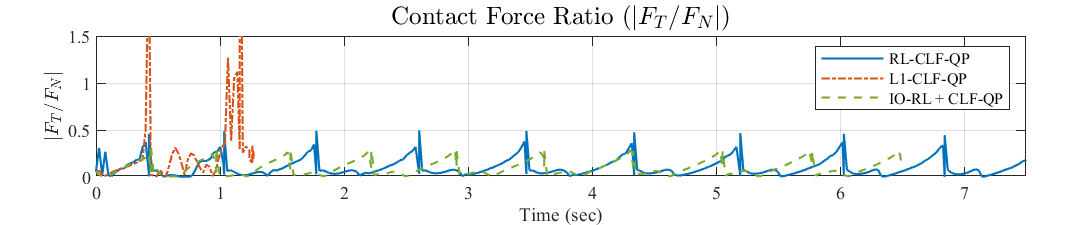}
\caption{Tracking error (top), its derivative (middle), and contact force ratio (bottom) of the three CLF-based controllers, simulated for ten walking steps with the additional torso weight 32kg (this amounts to the weight of RABBIT, i.e. 100\% additional weight).}\label{fig:clf-torso-results}
\vspace{-5pt}
\end{figure}

Displayed in Fig. \ref{fig:clf-torso-results} is the plot of tracking error and contact force ratio of the three controllers when, instead of the mass-inertia-scaling, an additional torso weight of 32kg (100\% of the robot mass) is introduced. It is notable that both the RL-CLF-QP and IO-RL + CLF-QP manage to adapt to this uncertainty, which has not been faced during the training. Furthermore, the RL-CLF-QP manages to stabilize the walking gait with an additional torso weight of up to 72kg (225\% of robot mass). On the other hand, IO-RL + CLF-QP manages to adapt to additional weights up to 53kg (166\% of robot mass).

\subsection{Simulation 2: Bipedal Walking on Stepping Stones with One Step Preview}

\begin{figure}
\centering
\includegraphics[width=0.28\textwidth]{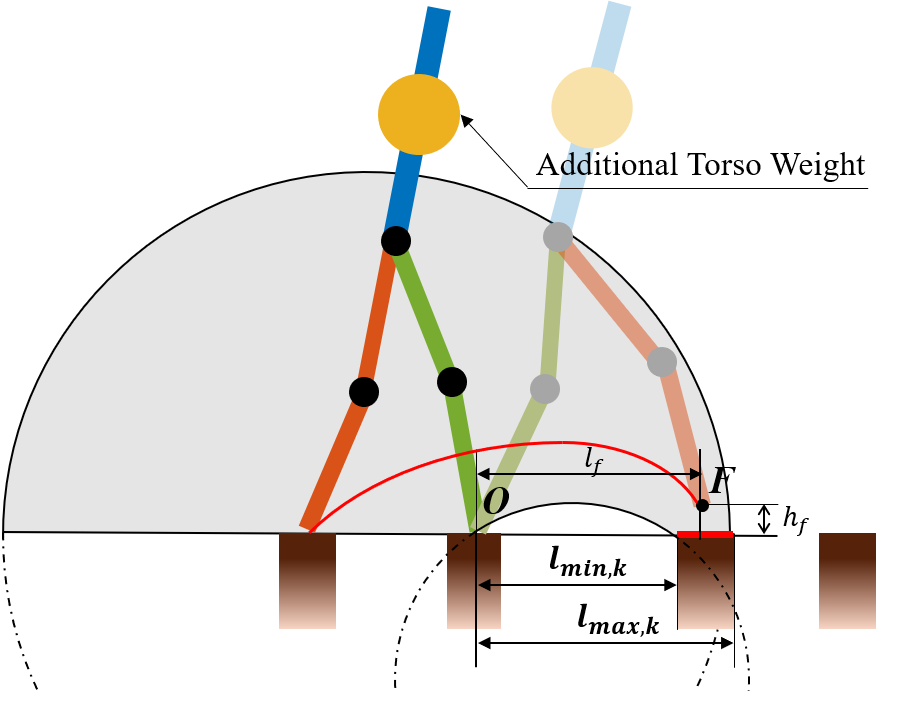}
\caption{Safety Constraint: In order to guarantee the swing foot lands on the stepping stone, we use two CBFs to ensure the swing foot position $F$ is within the grey area during the entire walking step.}\label{fig:stepping-stone}
\vspace{-10pt}
\end{figure}

We now evaluate the full RL-CBF-CLF-QP method with the safety-critical constraint of walking on stepping stones and the inclusion of friction constraints, which are dependent on the dynamics.
In this simulation scenario, for each step the robot faces a random placement of a stepping stone. Therefore, when the swing foot hits the ground at the end of the step, we want the step length to be within a specific range:
\begin{equation}
    l_{min, k} \leq l_{k} \leq l_{max, k},
\label{eq:steprange}
\end{equation}
\noindent where k indicates the step index.
Two position-constraints-based second order ECBFs parameterized by $l_{min, k},\ l_{max, k}$ that are a sufficient condition for \eqref{eq:steprange} are devised by \citet{nguyen2016optimal}. Basically these constraints imply that the swing foot position ($F$ in Fig. \ref{fig:stepping-stone}) needs to stay within the grey area. Note that $l_{min, k},\ l_{max, k}$ change for every step.

We also include contact force constraints in the RL-CBF-CLF-QP as control-affine dynamic constraints, following the procedure of Subsection \ref{subsec:RL-CBF-linear-constraint}. These are important since the original CBF-CLF-QP violates the friction cone and the unilateral normal force constraints repeatedly.

The robot is trained to walk on randomly spaced stepping stones, of which $l_{min}$ is sampled from a normal distribution $\mathcal{N}(0.35\text{m},0.02\text{m})$, truncated at $2.5\sigma$.
$l_{max}$ is set as $l_{min} + 0.05\text{m}$.

\begin{figure}
\centering
\includegraphics[width=0.5\textwidth]{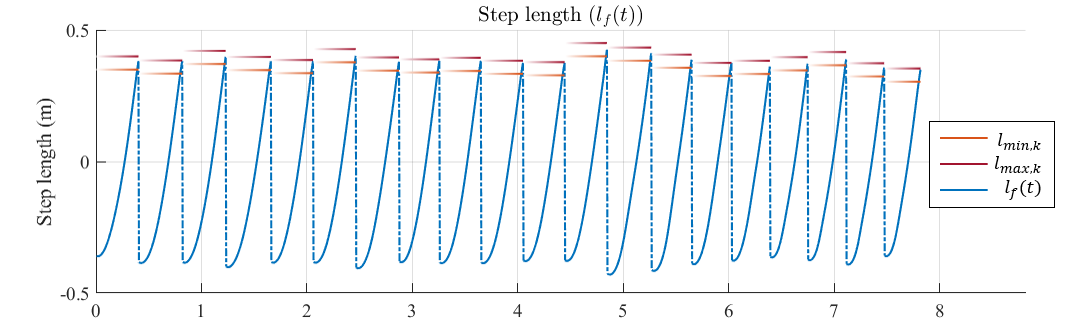}
\includegraphics[width=0.5\textwidth]{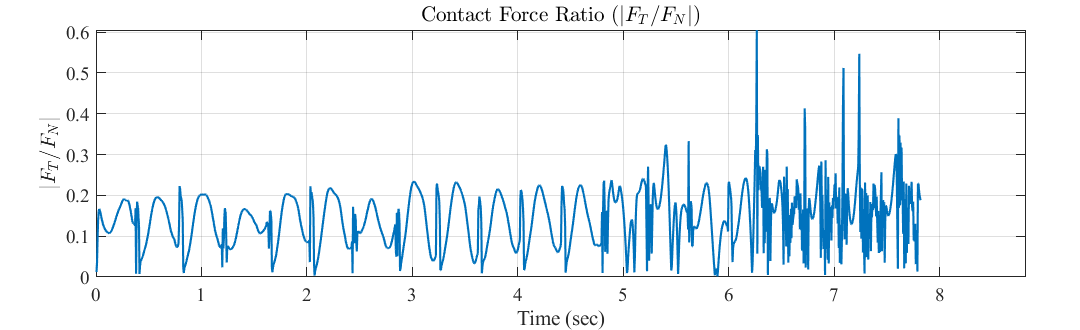}
\caption{Results of the simulation of 20 steps of walking on stepping stones, where the robot's mass and inertia values are scaled by a factor of 2. (Top) History of swing foot position $l_f$ for each step, with the stepping stone constraints $l_{min},\ l_{max}$. (Bottom) History of tangential-normal contact force ratio that satisfies to stay below $|F_T/F_N|\leq k_{f} = 0.8$.}\label{fig:rl-cbf-clf-results}
\vspace{-10pt}
\end{figure}

Fig.~\ref{fig:rl-cbf-clf-results} shows the result of the evaluation, where the robot walks on 20 randomly spaced stepping stones. We can check that the foot placement is always on the stepping stones. Also, it is verified that the contact force never exceeds the friction limit. Note that the sample distribution of $l_{min}$ here is same as during training.

Whereas our RL-CBF-CLF-QP method performs well, we have also tested the nominal model-based CBF-CLF-QP method on this simulation for comparison. The CBF-CLF-QP is also solved together with the friction constraints. However, it violates the step length safety constraints after an average of 5.6$\pm$4.64 steps. This value is obtained from 10 random executions of 20 steps simulation. 

Finally, for the case of having an additional torso weight applied to the original unscaled plant, RL-CBF-CLF-QP still manages to stay within the safety and friction constraints when the weight is in the range of [43kg, 72kg] (134-225\% of robot mass).

\section{Discussion}
\label{sec:discussion}

In Sections \ref{sec:application} and \ref{sec:results}, we have demonstrated that our method can compensate well for the trained model uncertainty and that it shows some robustness to the introduction of additional uncertainty during evaluation. It is important to note that our method is not restricted to mass and inertia scaling uncertainties, rather they have been used as illustrative examples for this paper. We have additionally tested our framework for other uncertainties: a simplified model of joint friction (assuming that joint friction reduces motor power by a 15\%, value taken from \citet{chevallereau2003rabbit}), joint damping (up to 1 $(rad/s^2)/(rad/s)$) and bending of links (up to 5\% of their length) obtaining successful results.

However, a primary drawback of our approach is that we need the designed nominal controller to not rapidly fail on the uncertain system before RL can learn the uncertainty.
This may not always be possible depending on the level of uncertainty. Following this same reasoning, for high levels of uncertainty the CLF designed for the nominal model may not be a CLF for the true plant, in which case our assumption would not hold and the method would fail. There is ongoing research on designing CLFs for systems with uncertain dynamics
\cite{richards18lyapunov,Umlauft18clf} that could be used to solve this issue, since our method is not restricted to any specific CLF. 

An illustration of the aforementioned limitation is that we have also tested our framework for mass-inertia uncertainty scales of 0.7 and 0.5. For the case of scale=0.7, our framework produces a stabilizing controller that respects safety and friction constraints for indefinitely long periods of walking, whereas the nominal model-based controller fails after just one step. In contrast, for the scale of 0.5, the nominal controller fails after just 0.06 seconds, which makes the training a lot more challenging and our framework fails.

Another limitation is that the measurements of $\dot{V}_{\varepsilon}$ and $B^{(r_b)}$ obtained from numerical differentiation could be noisy in experiments. However, a similar method is proved to be effective in real experiments in \citet{westenbroek2019feedback}, where an estimate of the output acceleration is computed by numerical differentiation, which is used to train the RL agent.

In this paper we specifically use RL to learn the uncertainty terms since in this way we can gradually enhance the quadratic program's feasibility and performance while learning the safety constraints, increasingly exploring the state space of our interest.
Moreover, RL allows us to unify the learning processes of uncertainty terms in multiple safety constraints to a single process. However, there are several works tackling similar problems with other learning methods, such as supervised learning \cite{taylor2019episodic,taylor2019learning}, and deciding which is the best approach is still an open question that might depend on the specific properties of the platform used for testing. We plan to address this in the future, adapting our safety-critical control framework to other learning methods.

Finally, it must be noted that feasibility of a CBF-CLF-QP with additional constraints, such as friction, is not guaranteed in general. However, using the trained RL-CBF-CLF-QP model, we observe that the feasibility drastically improves compared to the nominal CBF-CLF-QP.

\section{Conclusion}
\label{sec:conclusion}
We have addressed the issue of model uncertainty in safety-critical control with an RL-based data-driven approach. We have presented a formal analysis of uncertainty terms in CBF and CLF constraints, in addition to other dynamic constraints. Our framework includes two core components: 1) an RL agent which learns to minimize the effect of model uncertainty in the aforementioned safety constraints, and 2) the formulation of the RL-CBF-CLF-QP problem that solves online for the safety-critical control input. The proposed framework is tested on RABBIT, an underactuated nonlinear bipedal robot. We demonstrate walking on randomly spaced stepping stones with one step preview under high model uncertainty.

\bibliographystyle{plainnat}
\balance
\bibliography{references}

\end{document}